\documentclass[showpacs,aps,preprintnumbers,prb,amsmath,amssymb,twocolumn,floatfix]{revtex4-1}
\usepackage{graphicx}
\usepackage{color}
\DeclareGraphicsExtensions{.pdf,.eps,.png,.jpg,.mps}
\begin{document}
%%%%%%%%%%%%%%%
%\draft

\title{Superconductivity in the presence of disorder in skutterudite-related La$_3$Co$_4$Sn$_{13}$ and La$_3$Ru$_4$Sn$_{13}$ compounds; electrical transport and magnetic studies}
\author{A.~\'{S}lebarski$^{\star}$, M.~M.~Ma\'ska$^{\star}$, M.~Fija\l kowski$^{\star}$, C. A. McElroy$^{\dag}$,  and M.~B.~Maple$^{\dag}$}
\affiliation{
$^{\star}$Institute of Physics,
University of Silesia, 40-007 Katowice, Poland\\
$^{\dag}$Department of Physics, University of California, San Diego, La Jolla,
California 92093, USA
}
\begin{abstract}

La$_3$Co$_4$Sn$_{13}$ and La$_3$Ru$_4$Sn$_{13}$ were categorized as BCS superconductors. In a plot of the critical field $H_{c2}$ vs $T$, La$_3$Ru$_4$Sn$_{13}$ displays a second superconducting phase at the higher critical temperature $T_c^{\star}$, characteristic of inhomogeneous superconductors, while La$_3$Co$_4$Sn$_{13}$ shows bulk superconductivity below $T_c$.  
We observe a decrease in critical temperatures with external pressure and magnetic field for both compounds with $\frac{dT_c^{\star}}{dP} > \frac{dT_c}{dP}$. The pressure dependences of $T_c$ are interpreted according to the McMillan theory and understood to be a consequence of lattice stiffening. 
The investigation of the superconducting state of La$_3$Co$_x$Ru$_{4-x}$Sn$_{13}$ shows a $T_c^{\star}$ that is larger then $T_c$ for $x<4$. This unique and unexpected observation is discussed as a result of the local disorder and/or the effect of chemical pressure when Ru atoms are partially replaced by smaller Co atoms.

\end{abstract}


\maketitle

\section{Introduction}

The effect of atomic disorder on the electronic properties of correlated electron systems, particularly those close to a quantum critical point (QCP) \cite{JPCM96}  has been a topic of active research. In the critical regime, the system is at the threshold of an instability and even weak perturbations, e.g., disorder can cause significant effects by changing the nature of the quantum macro state. In these disordered systems, a rather large residual resistivity $\rho_0=\rho(T \rightarrow 0)$ is often encountered, even for single crystals, which means that even weak disorder is influential. As was argued theoretically \cite{Spalek2002}, such a drastic influence is possible because the band width of a few eV and an effective Hubbard interaction $U$ of the same order of magnitude result in a much more subtle energy balance that atomic disorder can disturb more easily. 
Therefore, investigations of atomic scale disorder in the form of defects and vacancies, granularity, and the effective increase in disorder by doping have received renewed attention in recent times particularly because of observations of novel phenomena in strongly correlated materials. 
The Kondo insulators are an example of thermoelectric materials where the defects lead to a high value of figure-of-merit
$ZT = S^2\sigma T/(\kappa_e + \kappa_L)$, 
where $S$ is the Seebeck coefficient, $\sigma$ is the
electrical conductivity, $\kappa_e$ is the electronic thermal conductivity,
and $\kappa_L$ is the lattice contribution to the thermal conductivity \cite{Zhang2011} due to the reduction of the lattice contribution to the thermal conductivity. 
The effect of disorder on superconducting properties has inspired a great deal of research, with
the discovery of unconventional superconductivity in heavy fermion compounds \cite{Steglich} and associated quantum  critical behavior. 
In many superconductors, the critical temperature $T_c$ decreases with increasing disorder and sufficiently strong disorder can, in fact, destroy superconductivity. As this disorder driven transition from a superconducting to a non-superconducting ground state occurs, the localization effects become so strong that often an insulating material results 
(at $T=0$ this is known as a quantum phase transition). This transition is referred to as a superconductor-insulator transition \cite{Goldman1998}. 
There are known strongly correlated superconductors that show evidence of nanoscale disorder, meaning that the sample 
exhibits electronic inhomogeneity over the length scale of the coherence length. Such substantial nanoscale electronic disorder is characteristic of Bi$_2$Sr$_2$CaCu$_2$O$_{8+x}$  high-$T_c$ 
materials, as well as PrOs$_4$Sb$_{12}$ \cite{Maple2002,Vollmer2003}, CePt$_3$Si \cite{Kim05,Takeuchi07}, and CeIrIn$_5$ \cite{Bianchi01}. Our recent investigation of the filled cage superconductors La$_3M_4$Sn$_{13}$ with $M=$ Rh \cite{Slebarski2014a} and Ru \cite{Slebarski2014b} have shown evidence of two superconducting phases: an inhomogeneous superconducting state below  $T_c^{\star}$ and the superconducting phase at  $T_c<T_c^{\star}$. This anomaly was interpreted in the context of the presence of an inhomogeneous superconducting phase between $T_c$ and $T_c^{\star}$. In this work, we present a comprehensive thermodynamic and high-pressure electrical resistivity study on La$_3$Co$_x$Ru$_{4-x}$Sn$_{13}$ to explain the superconductivity in the presence of disorder. La$_3$Co$_4$Sn$_{13}$ clearly exhibits a homogeneous superconducting phase at $T_c$, while in contrast La$_3$Ru$_4$Sn$_{13}$ and its Co-alloys  show evidence of nanoscale inhomogeneity with the presence of $T_c^{\star}$. The impact of disorder on the ground state of superconducting materials  has played an important role in condensed matter physics over the years. We believe that our results contribute towards developing a broader understanding of the complex behavior in novel superconducting strongly correlated electron systems.

\section{Experimental details}

Polycrystalline La$_3$Co$_4$Sn$_{13}$ and La$_3$Ru$_4$Sn$_{13}$ samples were prepared by arc melting the constituent elements on a water cooled copper hearth in a high-purity argon atmosphere with an Al getter. 
The dilute La$_3$Co$_x$Ru$_{4-x}$Sn$_{13}$ alloys were prepared by diluting nominal compositions of the parent compounds. The samples were then 
annealed at 870 $^{o}$C for 2 weeks. All samples were carefully examined by x-ray diffraction analysis and found to have a cubic structure (space group $Pm3n$) \cite{Remeika80} and for $x=1$ and 3.5, the samples were single phase while for $x=2$ and 3, the alloys were a mixture of two phases.
%to be single phase with cubic structure (space group $Pm\bar{3}n$) \cite{Remeika80} only for  $x=1$ and $x=3.5$, the remaining alloys with $x=2$ and 3 were obtained as a mixture of two phases.  

Stoichiometry and homogeneity were verified by the microprobe technique (scanning microscope JSM-5410) and by XPS analysis. As an example, measurements of La$_3$Co$_4$Sn$_{13}$ showed a composition close to the nominal ratio 3:4:13 (i.e., 14.87:19.93:65.20 for La:Co:Sn). 
For the La$_3$Ru$_4$Sn$_{13}$ and La$_3$Co$_x$Ru$_{4-x}$Sn$_{13}$ alloys, the composition of the samples 
were also close to the nominal ratio 3:4:13 stochiometry, and in Table \ref{tab:Table1}, we present the results from measurements of La$_3$Ru$_{3}$CoSn$_{13}$ noted at different points of the surface.

\begin{table*}[h!]
\caption{ Atomic \% reflecting the stoichiometric ratios for La$_3$Ru$_{3}$CoSn$_{13}$ sample at different areas on the surface.}
\label{tab:Table1}
\begin{tabular}{c|ccccc}
\hline
element &    &  & stoichiometry in at. \%  & &  \\
&  assumed   &  & measured &            &  \\
\hline
La & 15 & 15.94  & 15.55 & 17.07 &~~~~~~~~ 14.16    \\
Ru & 15 & 13.97  & 12.45 & 12.75 &~~~~~~~~ 13.70           \\
Co &  5 & 4.72   & 6.62  & 3.62  &~~~~~~~~ 5.07            \\
Sn & 65 & 65.37  & 65.38 & 66.47 &~~~~~~~~ 67.07    \\

\hline
\end{tabular}
\end{table*}

Ambient pressure electrical resistivity $\rho$ was investigated by a conventional four-point ac technique using a Quantum Design Physical Properties Measurement System (PPMS). 
Electrical resistivity measurements under pressure were performed in a beryllium-copper, piston-cylinder clamped cell. A 1:1 mixture of $n$-pentane and isoamyl alcohol in a teflon capsule served as the pressure transmitting medium to ensure hydrostatic conditions during pressurization at room temperature. The local pressure in the sample chamber was inferred from the inductively determined, pressure-dependent superconducting critical temperature of a Sn ingot \cite{Smith69}.

Specific heat $C$ was measured in the temperature range $0.4-300$ K and in external magnetic fields up to 9 T using a Quantum Design PPMS platform. The dc magnetization $M$ and magnetic susceptibility $\chi$ results were obtained using a commercial (Quantum Design)superconducting quantum interference device magnetometer from 1.8 K to 300 K in magnetic fields up to 7 T.

\section{Results and discussion}

\subsection{Electric transport, magnetic properties, and specific heat of La$_3$Co$_x$Ru$_{4-x}$Sn$_{13}$ near the critical temperature $T_c$ or $T_c^{\star}$}

\begin{figure}[h!]
\includegraphics[width=0.48\textwidth]{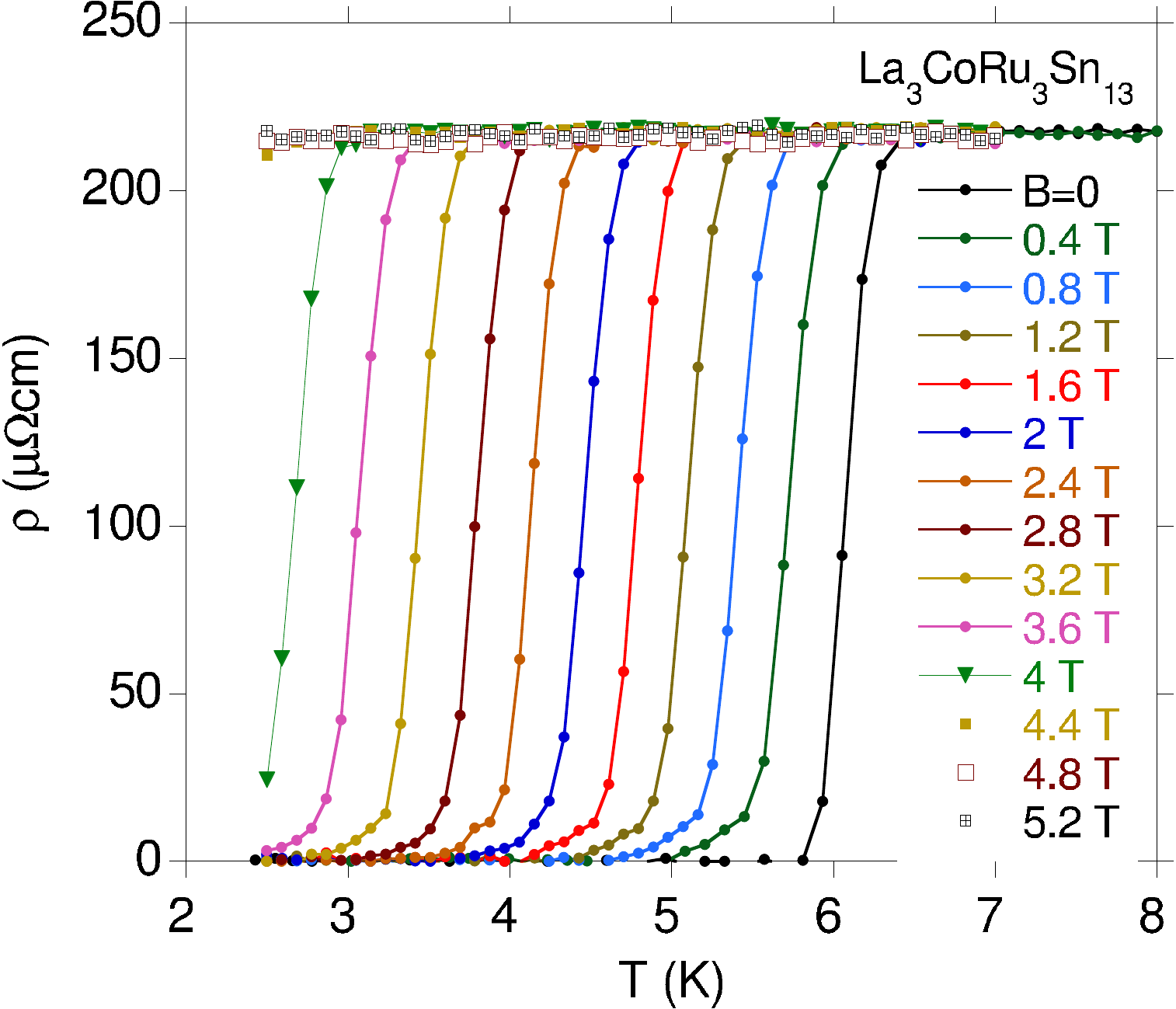}% Here is how to import EPS art
\caption{\label{fig:Fig1}
The temperature dependence of the resistivity $\rho$ of La$_3$CoRu$_{3}$Sn$_{13}$ at various externally applied magnetic fields, demonstrating the smooth suppression of $T_c^{\star}$.
}
\end{figure}

\begin{figure}[h!]
\includegraphics[width=0.48\textwidth]{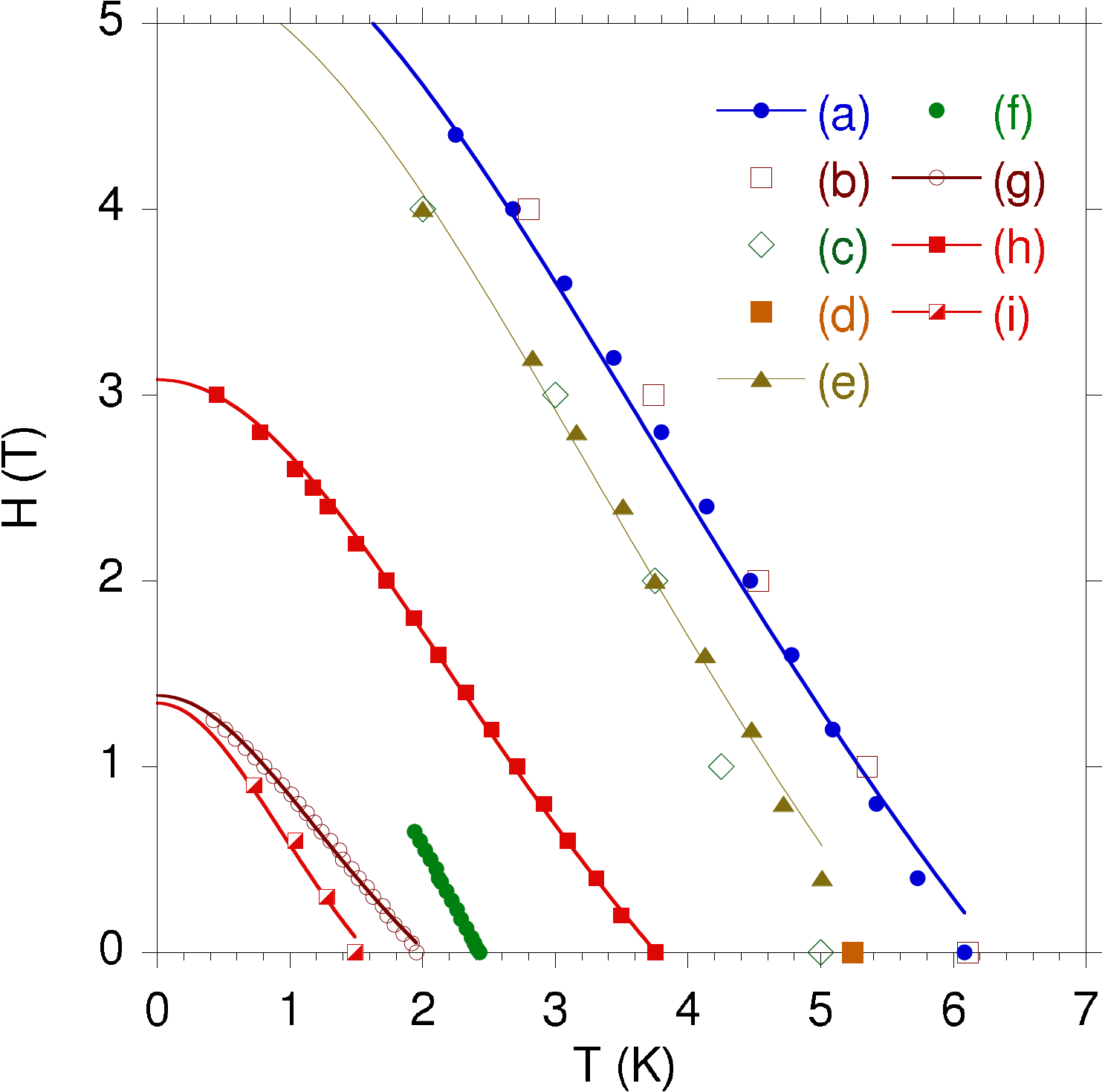}% Here is how to import EPS art
\caption{\label{fig:Fig2}
Temperature dependence of the upper critical fields $H_{c2}$  in the $H-T$ diagram for La$_3$Co$_x$Ru$_{4-x}$Sn$_{13}$. The solid lines represent a Ginzburg-Landau  fitting model for $H_{c2}(T)$. La$_3$CoRu$_{3}$Sn$_{13}$: Points $a$ represent $T_c$s obtained from resistivity under $H$ at 50\% decrease of the normal state $\rho$-value. Points $b$ show the temperatures where  anomalous behavior begins in $\Delta C/T$
at the high-temperature side of the specific heat peak. Points $c$ are attributed to temperature of the maxima in the best fits of $f(\Delta)$ to the experimental data $\Delta C(T)/T$, $d$ is the temperature of maximum in $\chi^{''}(T)$ at the magnetic field $B=0$, while $e$ show the temperatures where $\rho (T)\rightarrow 0$ (c.f.  Fig. \ref{fig:Fig1}). La$_3$Ru$_{4}$Sn$_{13}$: points $h$ represent $T_c^{\star}$ obtained from the $\rho$-data, while $i$ are $T_c$s from the specific heat data (Ref. \onlinecite{Slebarski2014b}.  La$_3$Co$_{4}$Sn$_{13}$: $g$ are $T_c$s from $\rho (T)$ at 50\% decrease of the normal state $\rho$-value (Ref. \onlinecite{Slebarski2014a}. $f$ are $T_c$s obtained from $C(T)$ vs $B$ for La$_3$Co$_{3.5}$Ru$_{0.5}$Sn$_{13}$ (c.f. Fig. \ref{fig:Fig3}).
}
\end{figure}

We performed a comprehensive thermodynamic and electrical resistivity study which reveal a homogeneous superconducting phase for La$_3$Co$_{4}$Sn$_{13}$, whereas for La$_3$Ru$_{4}$Sn$_{13}$ and the Co-substituted La$_3$CoRu$_{3}$Sn$_{13}$ samples there is evidence of two superconducting phases. 
Fig. \ref{fig:Fig1} shows results of resistivity measurements of La$_3$Ru$_{3}$CoSn$_{13}$ vs temperature $T$ in various megnetic fileds up to 5.2 T. Here we define the critical temperature at 50 \% of the normal state resistivity value. Similar $\rho (T)$ dependencies vs $B$ were presented for La$_3$Co$_{4}$Sn$_{13}$ and La$_3$Ru$_{4}$Sn$_{13}$ very recently (c.f. Refs. \onlinecite{Slebarski2014a,Slebarski2014b}). 
In Fig. \ref{fig:Fig2}, we show the $H-T$ phase diagram obtained for several investigated compounds and alloys of the system La$_3$Co$_{x}$Ru$_{4-x}$Sn$_{13}$, where $T_c$ is obtained from electrical resistivity under increasing magnetic fields (curves: $a$ for La$_3$CoRu$_{3}$Sn$_{13}$, $d$ for La$_3$Ru$_{4}$Sn$_{13}$, and $g$ for La$_3$Co$_{4}$Sn$_{13}$). 

The Ginzburg-Landau (GL) theory fits the data well and is shown in the $H-T$ plots in Fig.\ref{fig:Fig2}. The best fit of the equation $H_{c2}(T)=H_{c2}(0)\frac{1-t^2}{1+t^2}$, where $t=T/T_c$ gives the value of the upper critical field $H_{c2}(0)$ presented in Table \ref{tab:Table2}. Within the weak-coupling theory\cite{Schmidt77}  the upper critical field  through the relation $\mu_0 H_{c2}(0)=\Phi_0/2\pi \xi_0^2$ can be used to estimate the coherence length (where the flux quantum $\Phi_0=h/2e=2.068 \times 10^{-15}$ Tm$^2$), and these values are also listed in Table \ref{tab:Table2}. The table also displays the $H_{c2}(0)=0.693 T_c \frac{dH_c2}{dT}\mid_{T=T_c}$ \cite{Werthamer66,Helfland66} based on the results presented in  Fig. \ref{fig:Fig2}.

\begin{table*}[h!]
\caption{ Superconducting state quantities for La$_3$Co$_x$Ru$_{4-x}$Sn$_{13}$ near the bulk superconducting phase $T_c$ or the inhomogeneous phase below  $T_c^{\star}$.}
\label{tab:Table2}
\begin{tabular}{c|cccc}
\hline
La$_3$Co$_x$Ru$_{4-x}$Sn$_{13}$& $x=4$ & $x=3.5$ & $x=1$ & $x=0$  \\
\hline
$T_c$ (K)               & 1.95 & 2.41 & $\approx 5$      &  1.58 \\
$T_c^{\star}$ (K)       &      &      &  5.58 & 3.76 \\[1mm]
$\displaystyle\frac{dH_{c2}}{dT}\mid_{T=T_c}$ (T/K) & -0.904 & -1.34 & -1.3 ($T\rightarrow  T_c^{\star}$) &-1.33 ($T\rightarrow T_c$), -1 ($T\rightarrow T_c^{\star}$) \\ [3mm]
$H_{c2}(0)=0.693 T_c \displaystyle\frac{dH_{c2}}{dT}\mid_{T=T_c}$ (T) & 1.22 & 2.24 & 4.97 ($T\rightarrow T_c^{\star}$)& 1.45 ($T\rightarrow T_c$), 2.61($T\rightarrow T_c^{\star}$)  \\[2mm]
$H_{c2}(0)$ (T) & 1.38 & 3.05 & 5.28 &  1.34 ($T_c=0$), 3.08 ($T_c^{\star}=0$)\\
$\xi (0)$ (nm) & 16 & 11 & $\xi^{\star}(0)=8$ & 18, $\xi^{\star}(0)=9$ \\[1mm]
$\displaystyle\frac{\Delta C}{\gamma T_c}$ &1.5(5) & 1.7 & indefined & 1.6(1)\\[3mm]
$\displaystyle\frac{dT_c}{dP}$ (K/GPa) & 0.05 & -0.12 & & -0.03\\[3mm]
$\displaystyle\frac{dT_c^{\star}}{dP}$ (K/GPa) & & & -0.32 & -0.24 \\[2mm]
\hline
\end{tabular}
\end{table*}

\begin{figure}[h!]
\includegraphics[width=0.47\textwidth]{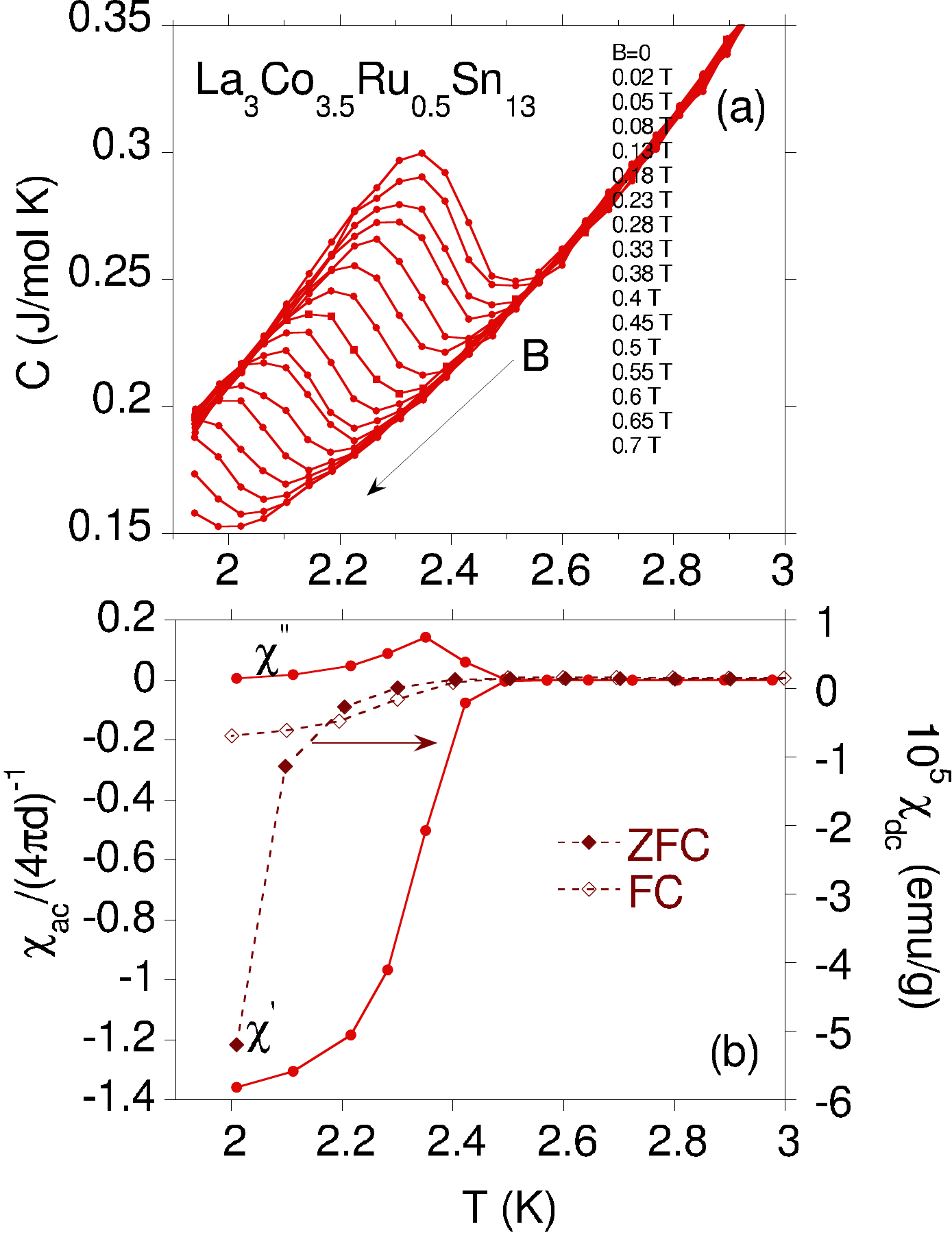}% Here is how to import EPS art
\caption{\label{fig:Fig3}
($a$) The temperature dependence of the specific heat $C(T)$ of La$_{3}$Co$_{3.5}$Ru$_{0.5}$Sn$_{13}$ at different magnetic fields $B$. ($b$) The ac magnetic susceptibility $\chi_{ac}$ at $B=12$ Gs divided by theoretical value of the full Meissner state $\chi^{'}=1/(4\pi d)$, and zero-field-cooled (ZFC) and field-cooled  (FC)  dc magnetic susceptibility in an applied field of $B=500$ Gs.
}
\end{figure}

Shown in Fig. \ref{fig:Fig3} is the specific heat $C$ plotted as $C$ vs $T$ at various magnetic fields (in panel $a$), and ac and dc magnetic susceptibility (panel $b$) for La$_{3}$Co$_{3.5}$Ru$_{0.5}$Sn$_{13}$. The heat capacity data for La$_{3}$Co$_{4}$Sn$_{13}$ (not shown  in Fig. \ref{fig:Fig3}, c.f. Ref. \onlinecite{Slebarski2014a})
indicates bulk superconductivity  below $T_c=1.95$ K in agreement with resistivity data, while La$_{3}$Co$_{3.5}$Ru$_{0.5}$Sn$_{13}$ shows 
a broad transition to the superconducting state with the same $T_c$  
from the resistivity and susceptibility data. 

\begin{figure}[h!]
\includegraphics[width=0.48\textwidth]{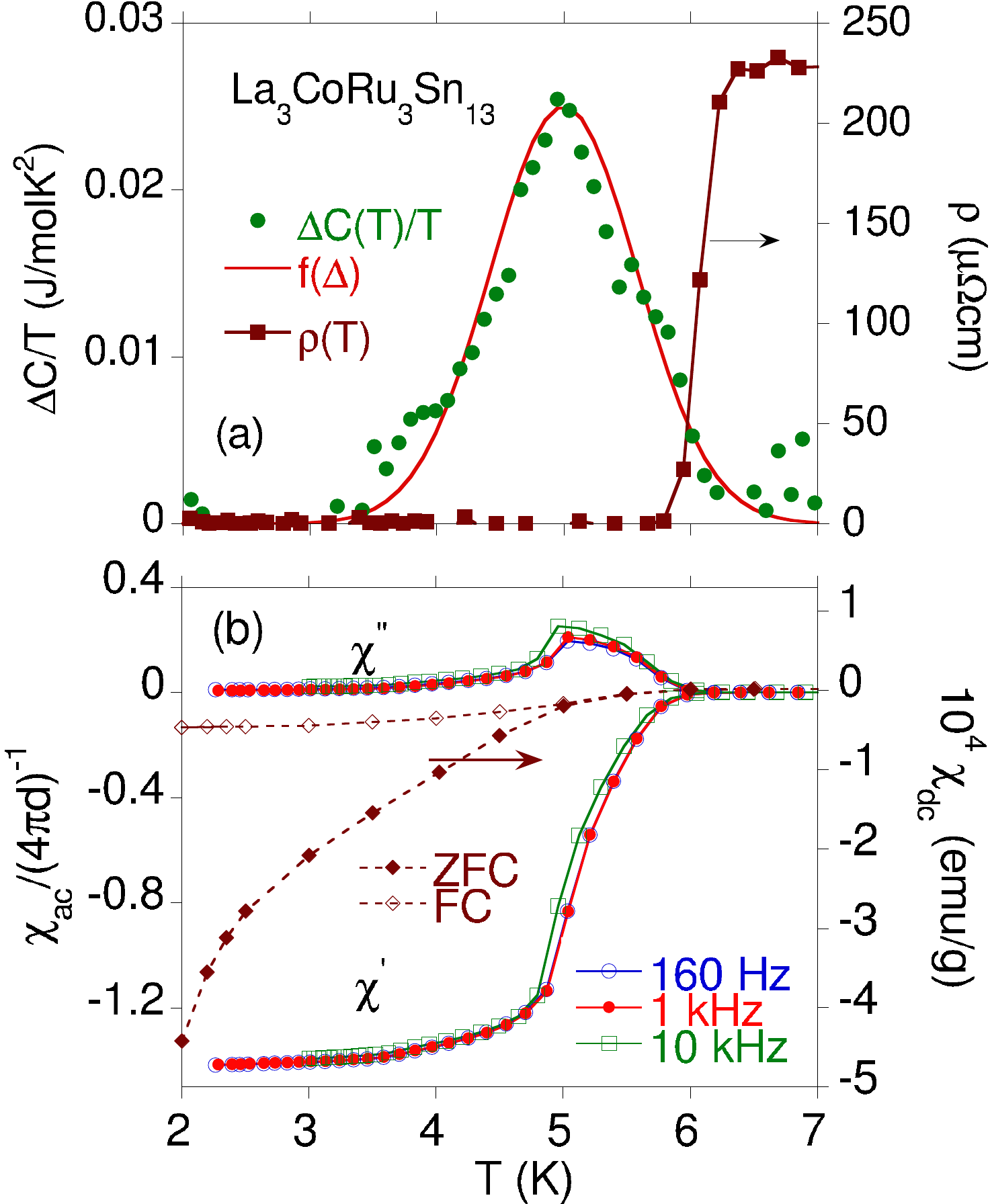}% Here is how to import EPS art
\caption{\label{fig:Fig4}
($a$) The temperature dependence of the specific heat $\Delta C(T)/T$ with a  Gaussian gap distribution fit $f(\Delta)$ and resistivity $\rho (T)$, both  at $B=0$ for La$_{3}$CoRu$_{3}$Sn$_{13}$.  For the sample under $B=0$,  $\Delta C(T)/T = C(T,B=0)/T - C(T,B=5 T)/T$, see Fig. \ref{fig:Fig5}. 
($b$) The ac magnetic susceptibility ($B=12$ Gs) $\chi^{'}$ and $\chi^{''}$ at different  frequencies, and ZFC and FC dc magnetic susceptibility ($B=500$ Gs).
}
\end{figure}

\begin{figure}[h!]
\includegraphics[width=0.48\textwidth]{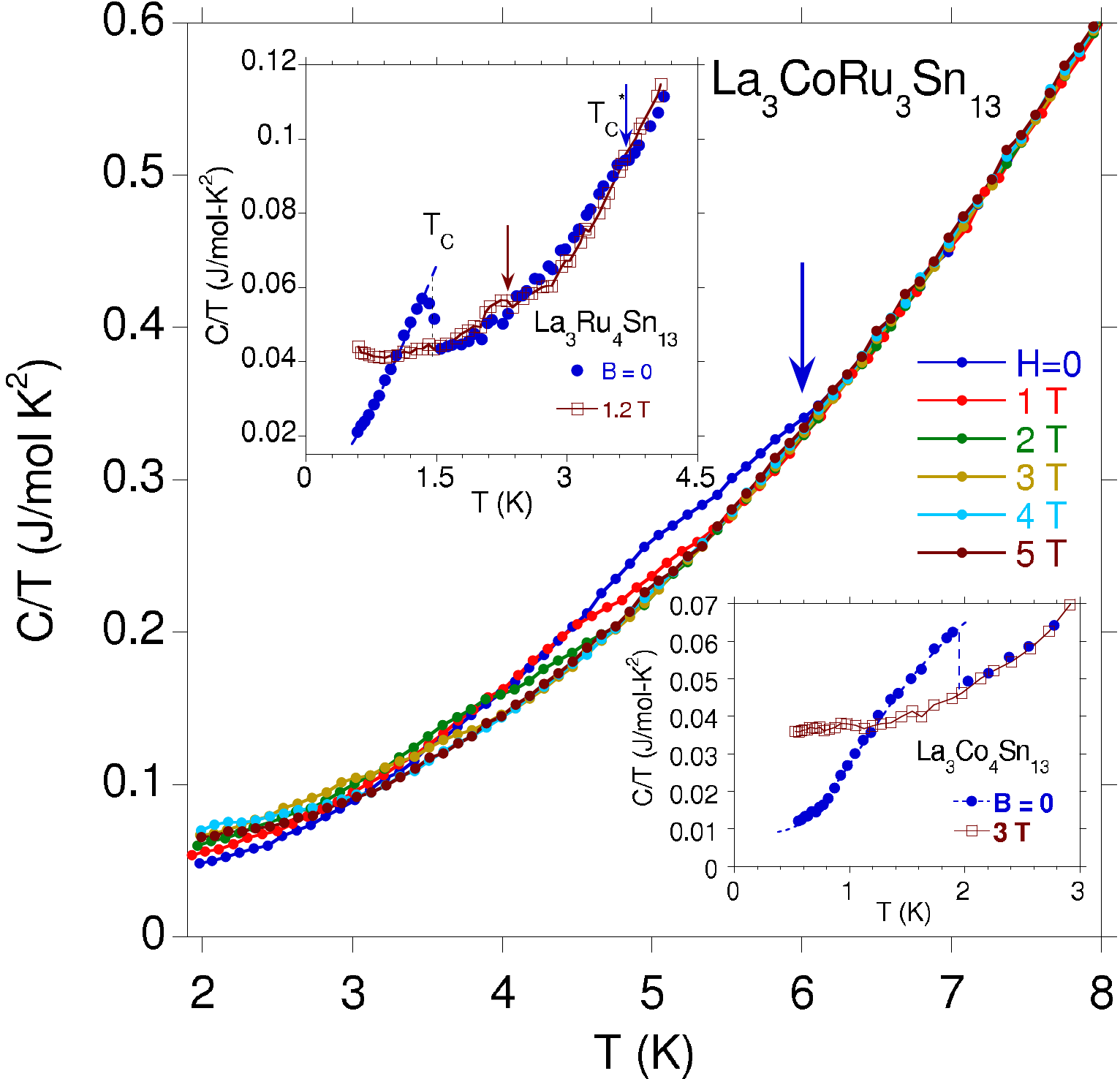}% Here is how to import EPS art
\caption{\label{fig:Fig5}
The specific heat $\Delta C(T)/T$ under various magnetic fields for La$_{3}$CoRu$_{3}$Sn$_{13}$. The arrow indicates the beginning of the superconductivity at $B=0$, the transitions under magnetic fields $B\neq0$ are similarly broad.
The insets display the $C/T$ data near $T_c$ for La$_3$Ru$_4$Sn$_{13}$ and La$_3$Co$_4$Sn$_{13}$. In case of La$_3$Ru$_4$Sn$_{13}$ the bulk effect at $T_c$ and inhomogeneous superconducting phase between $T_c^{\star}$ and $T_c$ are both shown. The dotted line is the best fit to the data for the expression $C(T)/T=\gamma +\beta T^2+A\exp(-\Delta(0)/k_BT)$.
}
\end{figure}

For La$_{3}$CoRu$_{3}$Sn$_{13}$, the superconductivity shown in $C/T$ data  (in Fig. \ref{fig:Fig4}$a$ and Fig. \ref{fig:Fig5}) also shows broad transition with the maximum in $\Delta C/T$ at $T_c\approx 5$ K spanning the maximum in $\chi^{''}$ in Fig. \ref{fig:Fig4}$b$. Under certain conditions, the ac losses in superconducting transition can exceed those of a normal metal, leading to a peak in $\chi^{''}$ vs $T$ \cite{Strongin63}. However, it was argued that a $\chi^{''}$ maximum can occur in  surface superconductors at sufficiently low frequencies, this is not the case in the ac magnetic susceptibility shown in Fig. \ref{fig:Fig3} \cite{comment1}.
The perfect diamagnetism of the full Meissner state $\chi^{'}=-1/(4\pi d)=9.55\times 10^{-3}$ emu/g for mass density $d=8.3$ g/cm$^3$ (Refs. \onlinecite{Thomas06,Mishra11}) is reached below the temperature of the maximum in $\chi^{''}$, it should also be noted that $\chi^{''}$ depends on the frequency of the  magnetic field, and that is characteristic of  magnetically inhomogeneous materials. 
We believe that
the superconductivity in La$_{3}$CoRu$_{3}$Sn$_{13}$ is
completely inhomogeneous superconductivity to explain the anomalies in the specific heat and $\chi^{''}$. 
Namely, we believe that the resistivity drop marks the onset of an inhomogeneous superconducting phase with spatial distribution of the magnitude of the superconducting gap, as a bulk property of the sample.
Since the drop of the resistivity at $T_c^{\star}$ is not accompanied by a change of $\chi^{''}$, the volume occupied by the inhomogeneous phase is too small to cancel out normal-state paramagnetic contributions. On the other hand, the superconducting regions must be  arranged as to form the necessary continuous paths reflected  in the resistivity measurements. 

Following Ref. \onlinecite{gabovich} we assume a simple Gaussian gap distribution 
\begin{equation}
f(\Delta)\propto \exp\left[-\frac{\left(\Delta-\Delta_0\right)^2}{2d}\right],
\end{equation}
where $\Delta_0$ and $d$ are treated as fitting parameters. The best fit of $f(\Delta)$ to the experimental data $\Delta C(T)/T$  gives the points $c$ in  Fig. \ref{fig:Fig2}, in good agreement with the points $e$ in Fig. \ref{fig:Fig2}. Points $d$ represents the temperature of the maximum in $\chi^{''}$.     
The behavior observed in this strongly disordered alloy is qualitatively different than that in La$_3$Ru$_{4}$Sn$_{13}$ \cite{Slebarski2014b}, or La$_3$Rh$_{4}$Sn$_{13}$ \cite{Slebarski2014a} with clear evidence for two superconducting phases: the {\it high temperature} inhomogeneous superconducting state below $T_c^{\star}$ and the second (bulk) superconducting phase  below $T_c$, where $T_c^{\star}>T_c$. We also note that the $C(T)/T$ data for La$_{3}$CoRu$_{3}$Sn$_{13}$ is not well approximated by  $C/T \sim \exp(-\Delta(0)/k_BT)$, while the bulk superconducting phases in both La$_{3}$Ru$_{4}$Sn$_{13}$ and La$_{3}$Co$_{4}$Sn$_{13}$ are well fit by this expression (c.f., Fig. \ref{fig:Fig5}). 
Excluding the case of La$_{3}$CoRu$_{3}$Sn$_{13}$, we found  $C(T)$ follows the behavior  described by the BCS theory in the weak-coupling limit, which indicates $s$-wave superconductivity. The BCS theory for $s$-wave superconductors provides a relation $\Delta C/(\gamma T_c)=1.43$ between the jump of the specific heat $\Delta C$ at the critical temperature $T_c$ and the normal metallic state contribution $\gamma$; the theoretical value $\Delta C/(\gamma T_c)=1.43$ is very close to the values presented in Table \ref{tab:Table2}.  

\begin{figure}[h!]
\includegraphics[width=0.48\textwidth]{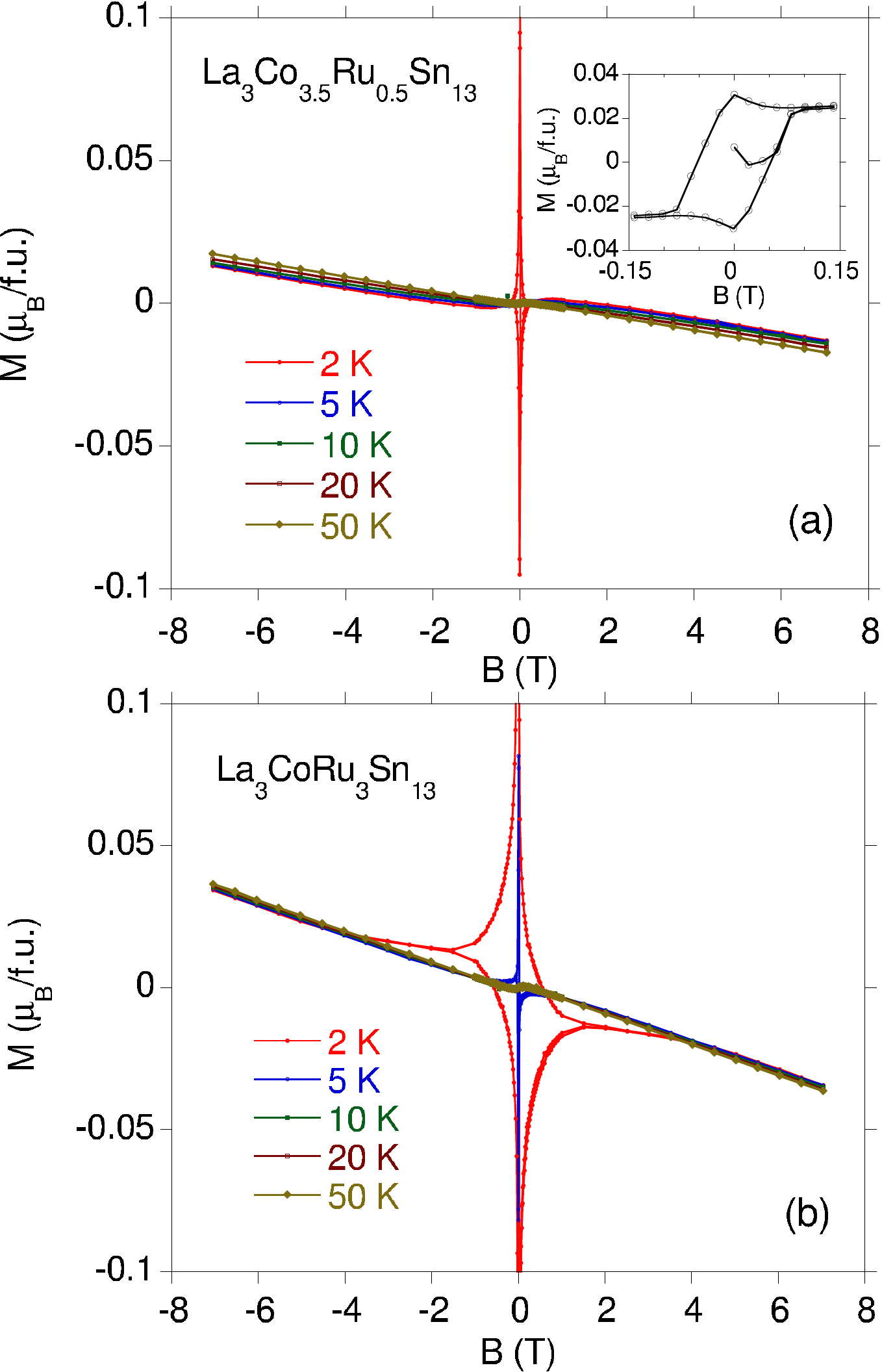}% Here is how to import EPS art
\caption{\label{fig:Fig6}
Magnetization $M$ per formula unit vs magnetic field at various temperatures. The inset shows a symmetric hysteresis loop at $T=1.8$ K for the superconducting state in La$_3$Co$_{4}$Sn$_{13}$. 
Panel ($a$) shows the data for La$_3$Co$_{3.5}$Ru$_{0.5}$Sn$_{13}$,  panel ($b$) displays $M$ for La$_{3}$CoRu$_{3}$Sn$_{13}$.
}
\end{figure}

In  Fig. \ref{fig:Fig6}, we show that the hysteresis loop in the superconducting state of La$_{3}$CoRu$_{3}$Sn$_{13}$ is about 3 T, while in the case of the remaining compounds, it is nearly an order of magnitude smaller. The broad hysteresis loop suggests strongly inhomogeneous material.

\begin{figure}[h!]
\includegraphics[width=0.48\textwidth]{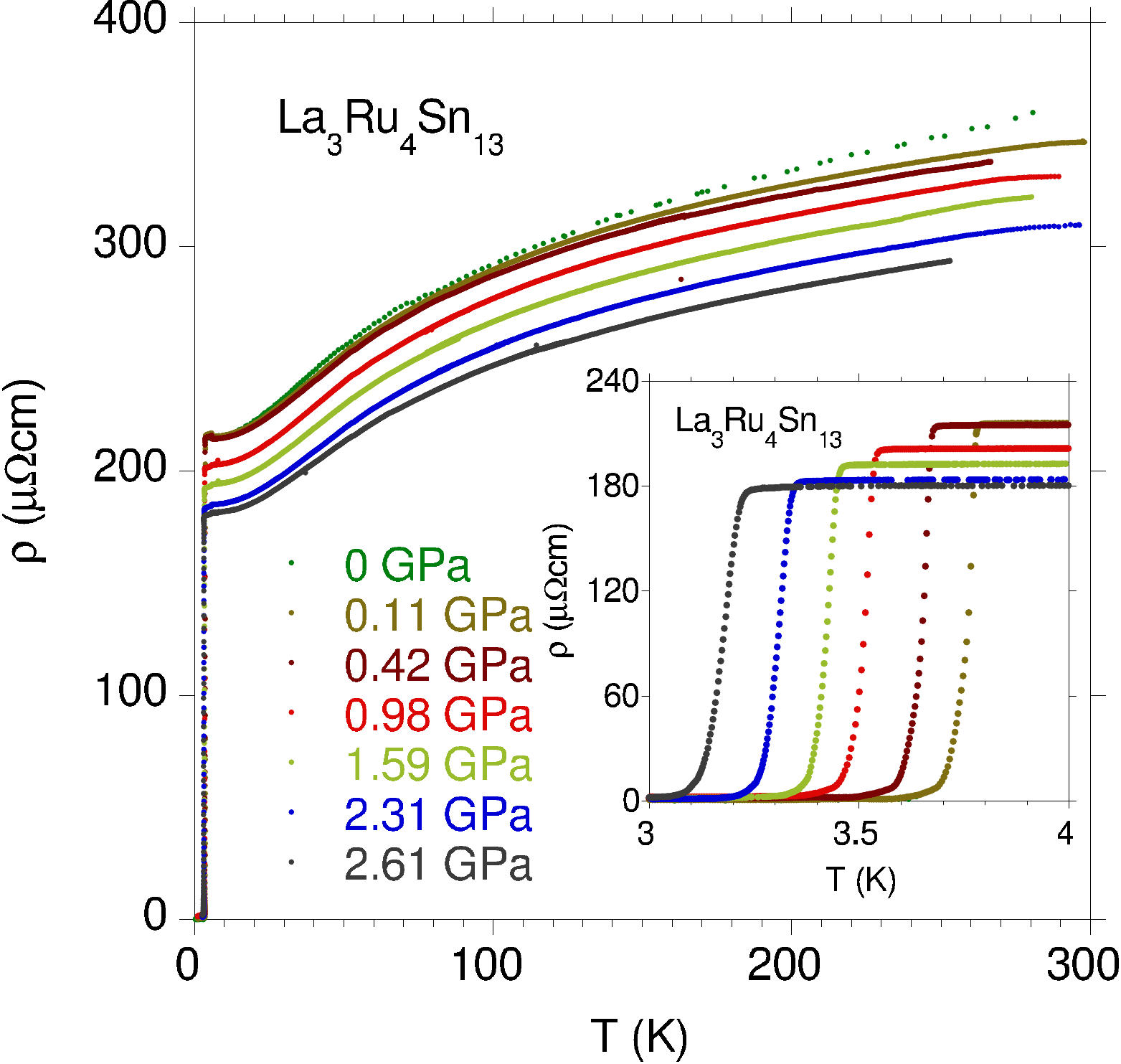}% Here is how to import EPS art
\caption{\label{fig:Fig7} Electrical resistivity $\rho$ of La$_{3}$Ru$_{4}$Sn$_{13}$ at various applied pressures. The inset displays the low temperature  details, showing the smooth suppression of $T_c$.
}
\end{figure}

\begin{figure}[h!]
\includegraphics[width=0.48\textwidth]{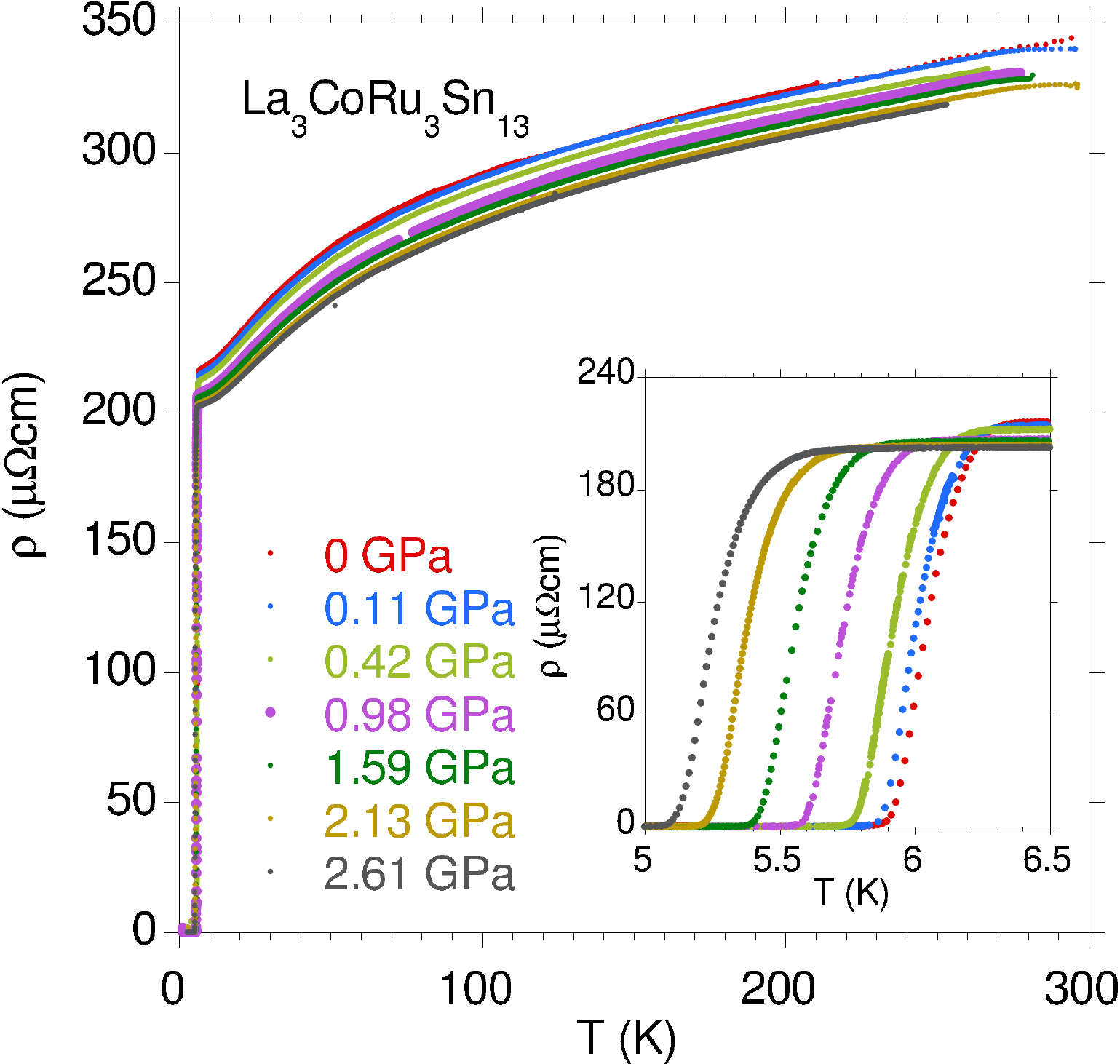}% Here is how to import EPS art
\caption{\label{fig:Fig8} Resistivity of La$_{3}$CoRu$_{3}$Sn$_{13}$ at different applied pressure. The inset displays the details.
}
\end{figure}

\begin{figure}[h!]
\includegraphics[width=0.48\textwidth]{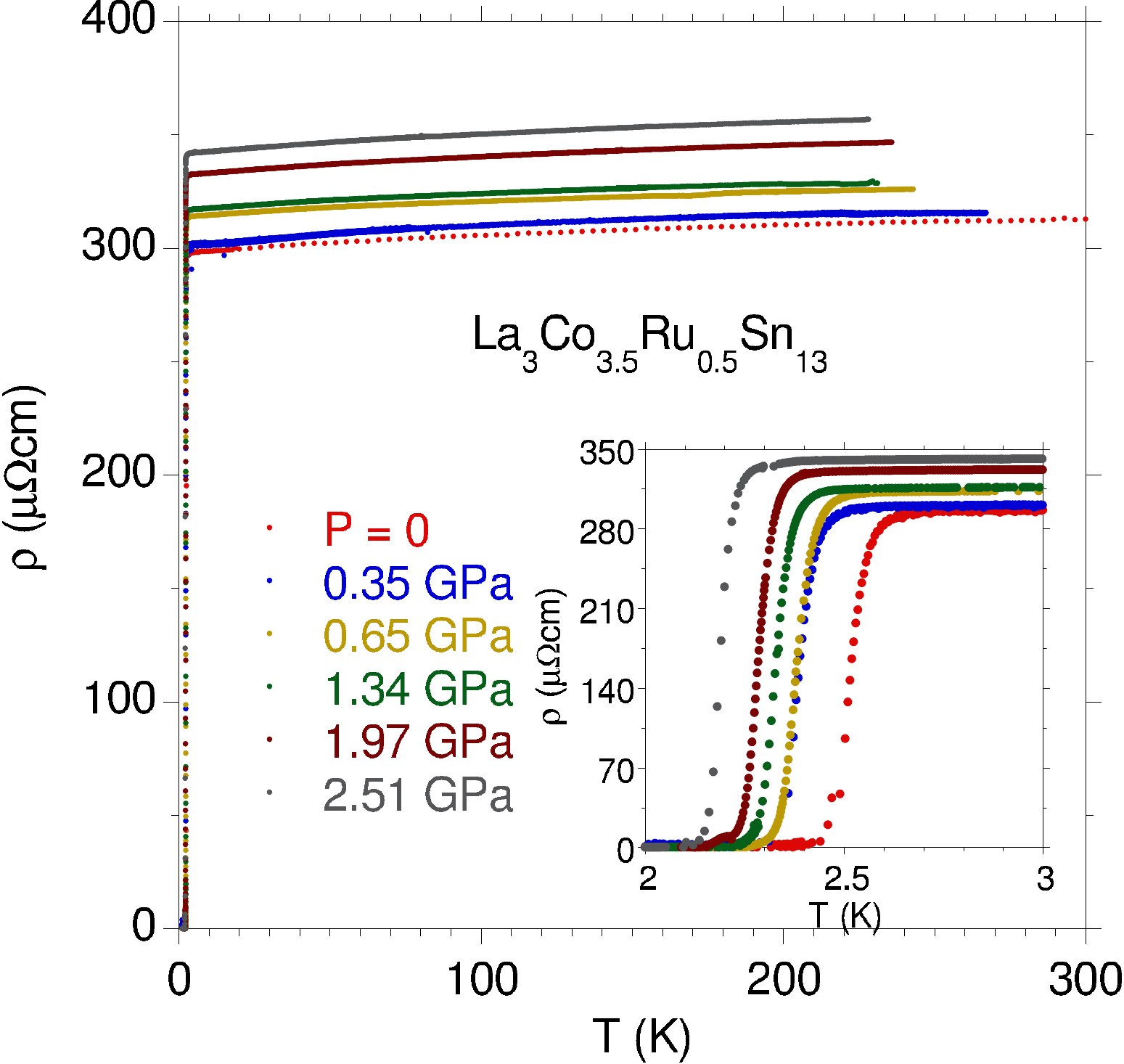}% Here is how to import EPS art
\caption{\label{fig:Fig9} Resistivity of La$_{3}$Co$_{3.5}$Ru$_{0.5}$Sn$_{13}$ at different applied pressure. The inset displays the details.
}
\end{figure}

\begin{figure}[h!]
\includegraphics[width=0.48\textwidth]{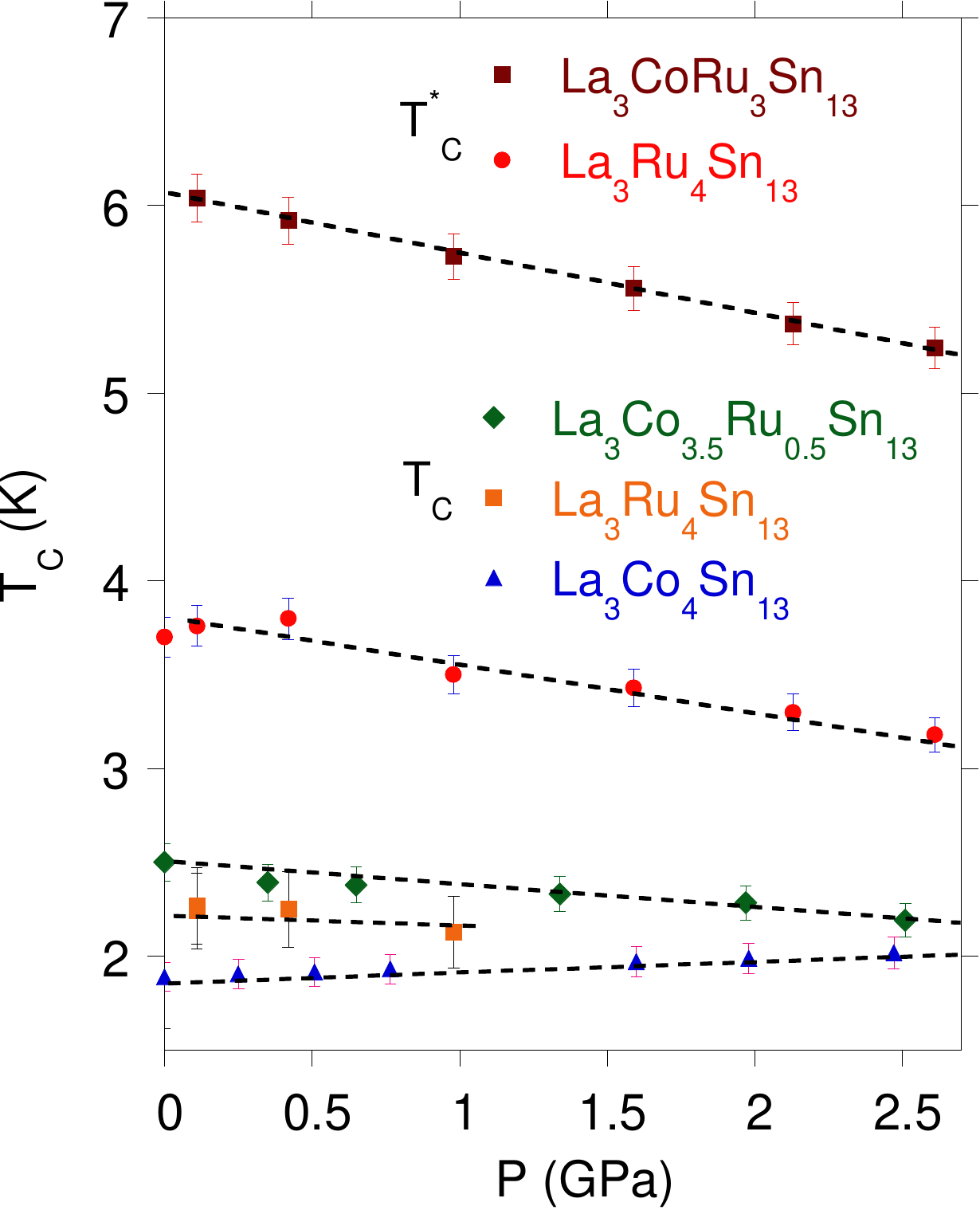}% Here is how to import EPS art
\caption{\label{fig:Fig10} Critical temperatures $T_c$ and $T_c^{\star}$ vs pressure $P$. The critical temperatures are obtained from resistivity under applied pressure at 50\% of the normal state value. For La$_{3}$Ru$_{4}$Sn$_{13}$, $T_c$ is estimated as a very weak change in $\rho (T)$ below $T_c^{\star}$.
}
\end{figure}

We expect that external pressure applied to strongly disordered materials should drive lattice instabilities from the compounds by varying the dominant parameters of the superconducting state, e.g., electronic density of states at the Fermi level. Most of the known superconductors show a decrease of $T_c$ with increasing applied pressure \cite{Olsen64}; however,  increasing pressure should also partially mitigate the inhomogeneity and stabilize the structural properties of the disordered system, and as a consequence, $T_c^{\star}$ is also expected to decrease with pressure. The evidence of this is shown in Figs. \ref{fig:Fig7}  - \ref{fig:Fig9} and summarized in Fig. \ref{fig:Fig10}. The observed increase of $T_c$ with pressure shown in Fig. \ref{fig:Fig10} for  La$_3$Co$_{4}$Sn$_{13}$ was recently discussed as a possible result of a subtle structural distortion below $T=140$ K \cite{Slebarski2014a}.

The pressure coefficients $\frac{dT_c}{dP}$ and $\frac{dT_c^{\star}}{dP}$ obtained from the respective $T_c$ vs $P$ data shown in Fig. \ref{fig:Fig10} are listed in Table \ref{tab:Table2}. The pressure coefficients of $T_c^{\star}$ are almost twice large as those of $T_c$, while for La$_3$Co$_4$Sn$_{13}$, the 
$\frac{dT_c}{dP}=0.05$ K/GPa is positive. 
The $P$-dependence of $T_c$ has been discussed according to the of Eliashberg theory of strong-coupling superconductivity \cite{Eliashberg61}. We employ the McMillan expression \cite{McMillan,Dynes72} 
\begin{equation}
T_c=\frac{\theta_\mathrm{D}}{1.45} \exp \left\{ \frac{-1.04(1+\lambda)}{\lambda - \mu^*(1+0.62\lambda)} \right\}, 
\end{equation}
which is a solution of the finite-temperature Eliashberg equations, to connect the value of $T_c$ with the electron-phonon coupling parameter $\lambda$, Debye temperature $\theta_D$ and the Coulomb repulsion $\mu^{\star}$
(the value of $\mu^*$ was chosen to be 0.1 as is typical for $s$ and $p$ band superconductors). This yields $\lambda\approx 0.4$ for $T_c$s and a larger $\lambda$ value $\sim 0.5$ for $T_c^{\star}$s. However, in the both superconducting states, relatively small $\lambda$ negates the strong coupling superconductivity. 
The coupling $\lambda$ is given by
\begin{equation}
\lambda=\frac{N(E_{\rm F})\langle I^2 \rangle}{M\langle \omega^2 \rangle},
\end{equation}
where $\langle I^2 \rangle$ is the square of the electronic matrix element of electron--phonon interactions averaged 
over the Fermi surface, $\langle \omega^2 \rangle$ is an average of the square of the phonon frequency, and $M$ is 
the atomic mass. Usually, $\mu^*$ and $\langle I^2 \rangle$ are  very weakly pressure dependent, so that the main
pressure effect on the transition temperature comes from $\theta_\mathrm{D}$ and $N(E_{\rm F})$ ($\langle \omega^2 \rangle$
depends on $\theta_\mathrm{D}$).
%
%Since the variation of the parameters $\theta_D$, $\mu^{\star}$, and $\lambda$ are volume (pressure) dependent, they are decided 
The pressure dependence of $\theta_D$ is given
by the Gr\"uneisen parameter $\gamma_G=-\frac{dln{\theta_D}}{dlnV}$, which represents the lattice stiffening. Using the McMillan expression it was found \cite{Shao2004} that $\gamma_G$ strongly determines the magnitude and sign of $\frac{dT_c}{dP}$. Our data suggest a larger $\gamma_G$ for the inhomogeneous superconducting state with respect to the bulk effect observed below $T_c$; in case of La$_3$Co$_4$Sn$_{13}$, the Gr\"uneisen parameter is expected to be smaller. It is also possible that in the case of inhomogeneous superconductivity, the pressure dependence of
the density of states at the Fermi level is more pronounced than in bulk superconductors, and may lead to a larger value of $\frac{dT_c^\star}{dP}$ than $\frac{dT_c}{dP}$.


\begin{figure}[h!]
\includegraphics[width=0.465\textwidth]{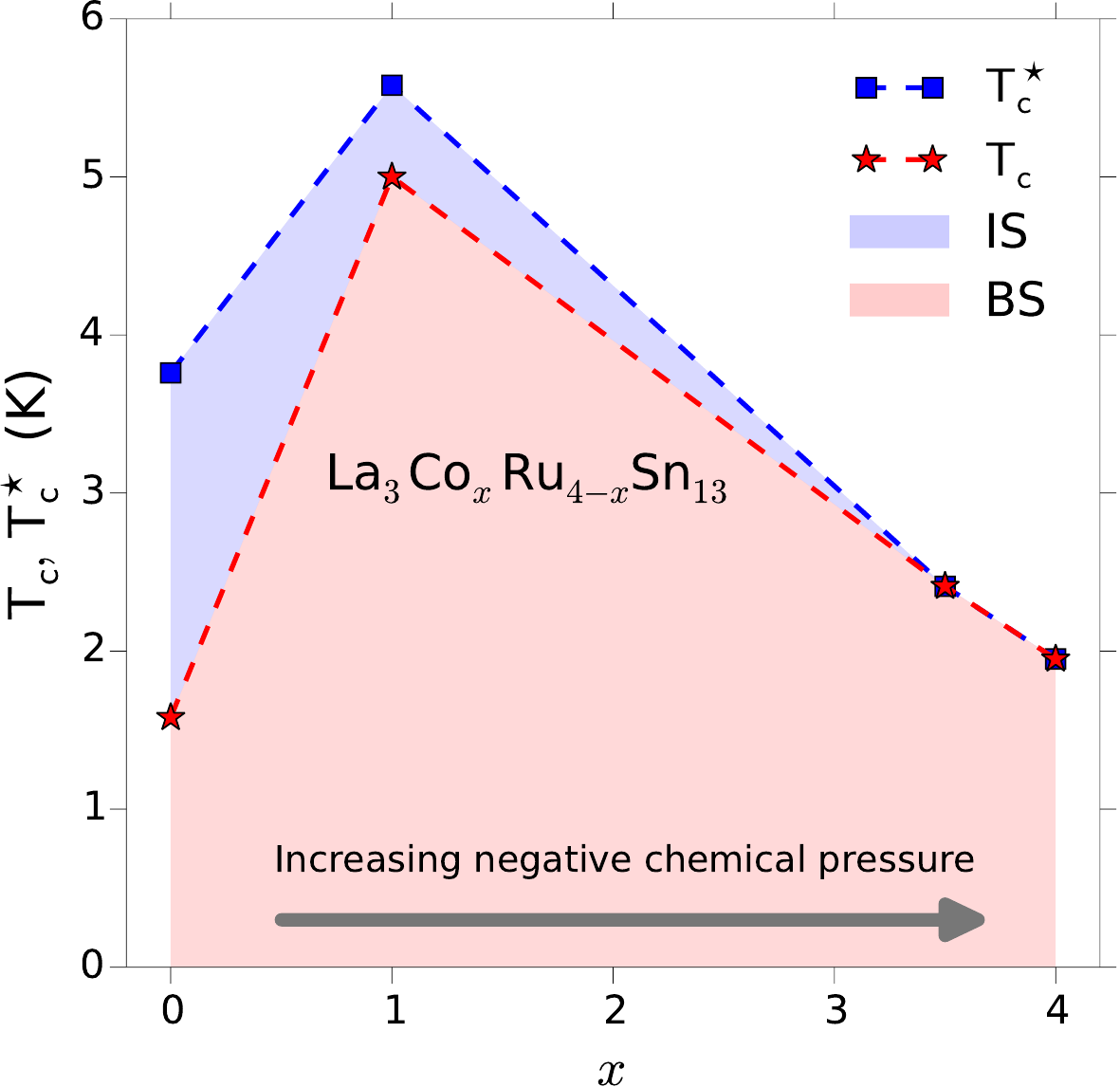}% Here is how to import EPS art
\caption{\label{fig:Fig11} Ambient pressure $T_c$ and $T_c^{\star}$ as a function of $x$ for La$_3$Co$_x$Ru$_{4-x}$Sn$_{13}$. Dashed lines are a guide to the eye. The blue area represents the inhomogeneous superconducting phase (IS), whereas the red area represents bulk superconductivity (BS).}
\end{figure}

Since the Co radius is smaller than that of Ru, increasing the amount of Co in the La$_3$Co$_x$Ru$_{4-x}$Sn$_{13}$ system leads to an effectively negative internal pressure. With $x$ increasing from 0 to 1, $T_c$ increases as well, but further, for $x$ going from 1 to  4, $T_c$ decreases almost linearly from 5 K to 1.95 K (see Fig.~\ref{fig:Fig11}). The dependence of $T_c^{\star}$ on the chemical pressure is consistent with the effects of  external pressure. With $x$ increasing from 0 to 1, $T_c^{\star}$ increases from about 3.76 K to 5.58 K, and can be interpreted as a continuation of the dependence on the external pressure for $P<0$. In this case, however, $T_c^\star$ is slightly less sensitive to the chemical pressure than $T_c$.

It is difficult to understand the abnormally large critical temperature  for La$_{3}$CoRu$_{3}$Sn$_{13}$, that is much larger than the $T_c^{\star}$ characterizing the inhomogeneous superconducting phase in of La$_{3}$Ru$_{4}$Sn$_{13}$. Typically,  disorder  strongly reduces the critical temperature due to the elevated impurity scattering; therefore, large value of $T_c^{\star}$ is surprising. 
It is, however, also possible to enhance superconductivity by local disorder \cite{Kivelson}. In this case, the superconductor may be inhomogeneous with lower and higher $T_c$ regions. Above $T_c^\star$, superconducting clusters appear, which at $T_c^\star$ form a network of continuous paths trough the entire sample. The random character of the Co substitution leads to a statistical (chaotic) distribution of these clusters. Despite the drop in the electrical resistivity, the fraction of the volume occupied by the superconducting state can still be small. This is a typical percolation scenario \cite{Kresin}. At a lower temperature $T_c$, the previously normal regions becomes superconducting and a macroscopic (bulk) superconducting state is formed. This is the transition that is seen in the specific heat and susceptibility measurements.

\section{concluding remarks}

In most of the known superconductors, the transition temperature $T_c$ decreases as a consequence of increased disorder. However, there are known examples of strongly correlated superconductors which show evidence of nanoscale disorder leading to an inhomogeneous superconducting state and, as a consequence, the critical temperature  $T_c^{\star}> T_c$. Both superconducting phases: the $T_c$-bulk phase and the $T_c^{\star}$ {\it high-temperature}  inhomogeneous  phase whose onset is  observed between $T_c^{\star}$ and $T_c$ are present in the skutterudite-related La$_3$Rh$_4$Sn$_{13}$ and La$_3$Ru$_4$Sn$_{13}$ compounds. In these compounds, we observed a decrease of the critical temperature with the application of  external pressure, however, the pressure coefficients $\frac{dT_c^{\star}}{dP}$ are nearly twice as large as their respective $\frac{dT_c}{dP}$ values. In the case of La$_3$Co$_4$Sn$_{13}$, $\frac{dT_c}{dP}$ is positive. The $P$-variations of $T_c$ were interpreted in the context of the Eliasberg theory and discussed as a consequence of the lattice stiffening. 
The results shown in this work should be of interest for understanding the $x$-dependent superconducting state of  La$_3$Co$_x$Ru$_{4-x}$Sn$_{13}$ where $T_c^{\star}$ is larger than $T_c^{\star}$ for La$_3$Ru$_4$Sn$_{13}$, (e.g., for La$_3$CoRu$_{3}$Sn$_{13}$ it is almost twice as large). This unique observation is not predicted  by the BCS theory and not observed in other chemically substituted superconductors. We suggest that local disorder is responsible for the increase in $T_c$ in strongly inhomogeneous regions in the sample and/or the effect of chemical pressure when La$_3$Ru$_{4}$Sn$_{13}$ is substituted with by Co. This
scenario should be verified theoretically. 

\section{acknowledgments}

The research was supported by National Science Centre (NCN) on the basis of Decision No. DEC-2012/07/B/ST3/03027. M.M.M. acknowledges support by NCN under grant DEC-2013/11/B/ST3/00824. 
High-pressure research at the University of California, San Diego, was supported by the National Nuclear Security Administration under the Stewardship Science Academic Alliance program through the U. S. Department of Energy under Grant Number DE-NA0001841. One of us (A.\'S.) is grateful for the hospitality at the University of California, San Diego (UCSD).

\newpage

\end{document}